\documentclass[]{spie}  
\usepackage[]{graphicx}

\title{The Gemini Observatory Fast Turnaround Program} 

\author{R. E. Mason\supit{a},  S. C\^{o}t\'{e}\supit{b}, M. Kissler-Patig\supit{a}, N. A. Levenson\supit{a}, A. Adamson\supit{a}, C. Emmanuel\supit{b,c} and D. Crabtree\supit{b}
\skiplinehalf
\supit{a}Gemini Observatory, Northern Operations Center, 670 N. A'ohoku Place, Hilo, HI 96720, USA; \\
\supit{b}NRC Herzberg Astrophysics, 5071 W. Saanich Road, Victoria, BC V8Z 3H7, Canada \\
\supit{c}University of Victoria, 3800 Finnerty Rd, Victoria, BC V8P 5C2, Canada \\
}

\authorinfo{Further author information: \\R.E.M.: E-mail: rmason@gemini.edu}

\begin{document} 
  \maketitle 

\begin{abstract}

Gemini's Fast Turnaround program is intended to greatly decrease the time from having an idea to acquiring the supporting data. The scheme will offer monthly proposal submission opportunities, and proposals will be reviewed by the principal investigators or co-investigators of other proposals submitted during the same round. 
Here, we set out the design of the system and outline the plan for its implementation, leading to the launch of a pilot program at Gemini North in January 2015.

\end{abstract}


\keywords{Gemini Fast Turnaround, Rapid Response, Observatory Operations, Proposal Assessment Mechanisms, Proposal Selection Processes, Novel Observing Modes, Distributed Peer Review}

\section{INTRODUCTION}
\label{intro}  

The standard system for obtaining time on most astronomical facilities involves a delay of many months, if not years, between conceiving an idea and acquiring the supporting data. Gemini currently receives many applications for Director's Discretionary Time (DDT) that do not really fit the DDT criteria of being urgent, high-impact, or risky, but which nonetheless reflect commendable enthusiasm, creativity and eagerness for data among the user community. By offering a good mechanism for receiving and assessing proposals and obtaining data within a matter of weeks,  Gemini could enable this enthusiasm to be quickly transformed into published results, benefitting both users and the observatory. We have encountered very positive reactions to this possibility and are working on making it a reality.

Under the ``Fast Turnaround'' (FT) program, astronomers will have the opportunity to submit proposals every month (as opposed to every six months in the standard scheme). The outcome of the proposal selection process will be known in a little over two weeks, and observations made ready for execution within a month of the proposal deadline. Three dedicated Fast Turnaround nights per month will be scheduled for the accepted programs, which will remain valid in this ``mini-queue'' for three months. A novel feature of the Fast Turnaround program is that principal investigators (PIs) of proposals will review {\em each other's} proposals, and their combined assessment of scientific merit will determine the final ranked list of programs from each proposal round.

Initially envisaged as a pilot program running for a few months and accounting for around 10\% of the time at one telescope, the Fast Turnaround scheme will occupy the opposite end of the scale to Gemini's new Large and Long Program system. While that program will accept annual proposals for projects requiring a large amount of time or extending over long periods of time, the Fast Turnaround program will allow users to acquire a few hours' worth of data to -- for example -- finish a thesis project, follow up an unexpected astronomical event or discovery, or test a technique to validate plans for a larger proposal. The regular, six-monthly time allocation committee (TAC) process will continue to exist for the foreseeable future, catering to those who may prefer not to use the Fast Turnaround program, with its fairly small amount of time, peer review system, and somewhat limited observing modes. DDT will remain the mechanism for obtaining time for unforeseen, potentially high-impact observations without these constraints.

In this paper we present the design of the Fast Turnaround program, describe the current status of the project, and outline the plan for its implementation. The approach is that the program should be well thought out and practical, while acknowledging that the ``final'' answer will come as we gain experience during the trial period. Close monitoring of the process, and good communication with our users, will be the key to refining it into a successful long-term program.

\section{PRINCIPLES}
\label{principles}

The Fast Turnaround program will adhere to the following top-level principles:

\begin{itemize}
\item The process has to lead to the production of good science 
       \begin{itemize}
       \item There must be no reduction in quality compared to the standard proposal process
       \item The scheme must be driven by users' scientific needs
       \end{itemize}
\item The process must maintain integrity     
       \begin{itemize}
       \item The review process must be fair, transparent, and no more susceptible to perceived or actual
abuse than the current standard TAC process
       \end{itemize}
\item  The system should be clearly distinct from the standard proposal process
        \begin{itemize}
        \item The turnaround should be rapid enough to clearly distinguish this from the regular proposal process
        \item The process should encourage rapid publication, and should not be simply an alternative way of accessing regular queue time
        \end{itemize}
\item The process should encourage community engagement with Gemini
        \begin{itemize}
        \item This should be something genuinely new and different
        \item The system needs to be well-communicated and user-friendly
        \end{itemize}
\item The process needs to allow creativity
        \begin{itemize}
        \item It should impose as few rules and restrictions as possible, as far as is compatible with the
other top-level principles
        \end{itemize}
\item It must be possible to scale the implementation
        \begin{itemize}
        \item If a pilot implementation shows demand for this mode, there should be no practical barriers to increasing the time available
        \end{itemize}
\item The process must be sustainable and compatible with Gemini's current operations
        \begin{itemize}
        \item We have to implement a process that Gemini can competently and consistently support
        \item We should be ambitious and creative, while realizing that it would be a mistake to promise
things we cannot deliver
        \end{itemize}
\item The process must allow partner shares to be monitored and respected
\end{itemize}

These principles, as well as the practicalities of adding a new observing mode to those currently offered by the observatory, were borne in mind while designing the program. 

\section{SYSTEM OVERVIEW}
\label{system}

\begin{figure}
   \begin{center}
   \begin{tabular}{c}
   \includegraphics[height=10cm]{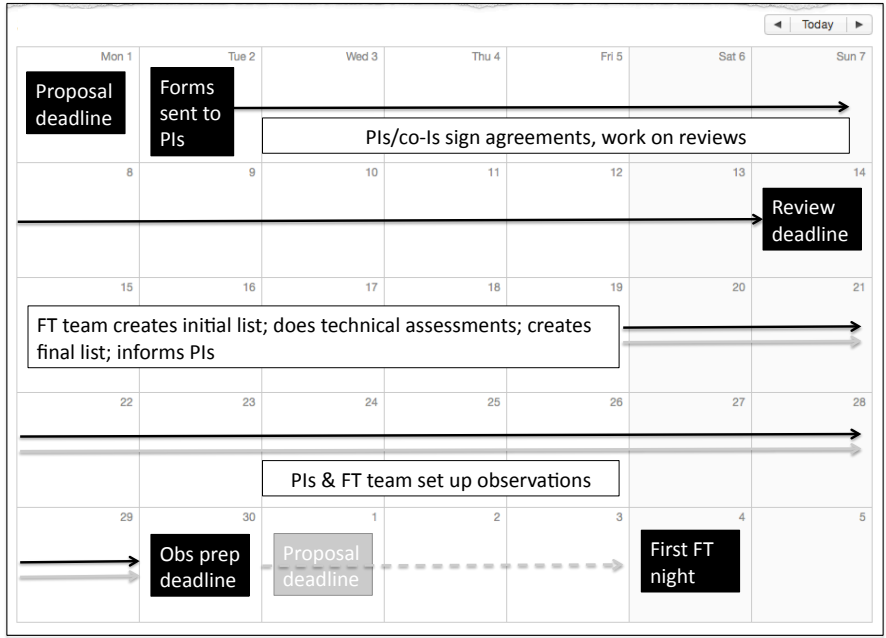}
   \end{tabular}
   \end{center}
   \caption{Example of a month in the life of the Fast Turnaround program. Solid boxes denote fixed events or deadlines, open boxes describe work done by PIs (black arrows) and/or the Fast turnaround team (gray arrows). The dates of the Fast Turnaround nights will be set at the start of each semester. \label{f1}}
   \end{figure}

The Fast Turnaround program will operate on a monthly cycle, illustrated in Figure \ref{f1}. Proposal deadlines will fall at the end of each month, and the proposals will be reviewed within two weeks of the deadline. Because of the likely need to review many proposals each month, the assessments will be carried out not by the usual TAC, but by the principal investigators (PIs) or co-investigators of other proposals submitted during the same round. Once a ranked list of proposals has been created from the assessors' individual grades, Gemini staff will create the final list of accepted proposals by taking into account the likely time available in the various weather condition bins. At this stage, less than three weeks after the deadline, PIs will be informed of the outcome of this proposal round. They will then work together with Gemini staff to ready their observations before the end of the current month, so that their program is ready to be observed by the time of the next scheduled Fast Turnaround night at the telescope, and FT observations will remain valid for three months. Each of these steps is described in more detail in the following sections.

\subsection{Proposal Submission}
\label{propose}



The monthly proposal deadlines will be advertised on the web, along with information about the available instrumentation, target visibility, expected number of hours available in various weather condition bins, etc. The existing Phase I Tool, with minor modifications, will be used for submission of Fast Turnaround proposals.

Target visibility criteria will be such that it must be possible to execute the proposed observations during the 3 months following that month's observation preparation deadline (i.e., 1-4 months after the proposal deadline). In the rapid-turnover spirit of the program, and to ensure that it is not simply a second way of gaining access to the regular queue, it will be required that the earliest targets be observable starting immediately after the observation preparation period (or as soon as the requested instrument is available). This will be one month after each proposal deadline. 

Some restrictions will be placed on the observing modes and instrument configurations that can be proposed for, at least during the initial trial period. For instance, it may be possible to accept multi-object spectroscopy observations, but only if they do not require pre-imaging, and with some limitations on the number of masks that can be accommodated. Support for ``non-standard'' observing modes (e.g. using the acquisition camera to observe an occultation) is not guaranteed and will depend on the likely availability of staff qualified to assist with that mode. Rather than set well-defined limits on what can be proposed for, we will give examples of ``difficult'' observations and encourage prospective users to contact us in advance of proposing for similar things. Our philosophy will to be to start simply and phase modes in (or out) as we gain experience with the program.

\subsection{Proposal Review}
\label{review}

The proposals will be reviewed and graded by the PIs or co-Is of other proposals submitted during that round. Submitting a proposal will commit the PI/co-I to review up to approximately 10 other proposals. After identifying conflicts of interest and agreeing to keep the proposals confidential and use them only for the purpose of the review, reviewers will be asked to grade the proposals and provide brief written reviews and  self-assessments of their knowledge of the subject area. Failure to submit a review within two weeks will result in the PI's own proposal being automatically removed from consideration. A dedicated Fast Turnaround Team of three or four staff scientists (of whom two will be on duty each month) will oversee this process, using software already developed at NRC-Herzberg, as well as gathering statistics and information to help evaluate and improve the program. This proposal assessment model, and the reasons for choosing it, are discussed in more detail in \S\ref{peer_review}.

\subsection{Observation Selection and Preparation}
\label{select}

At this stage, a ranked list of proposals will be available to the Fast Turnaround Team. Their job is then to construct the final set of accepted programs. They will not make any scientific judgments about the programs, although they will read and evaluate the reviews as part of the overall monitoring of the Fast Turnaround model. The team will need to determine: \\
\\
(1) Which of the highest-ranked programs will fit in the available time? \\
(2) Are there any technical reasons not to schedule any of those programs?

By answering question (1), we aim to avoid filling the Fast Turnaround nights with programs that are unlikely to be executed. Achieving a good balance of weather conditions will require some effort even though, by adopting distinct FT nights (\S\ref{exec}), the state of the regular queue does not have to be taken into account. To make things easier, we will require that the length of a proposal be no more than the time available in the most restrictive observing condition. These numbers would be publicized in the Fast Turnaround call for proposals (\S\ref{propose}). For instance, if we scheduled 3 nights/month, after $~$25\% weather loss, 23 hours would be left for science, of which 4.5 hours would be expected to be in the ``IQ20'' or ``excellent seeing'' bin. A program requiring IQ20/CC70/WVAny/SBAny (excellent seeing, thin cloud, unrestricted sky background and water vapor) could therefore ask for no more than 4.5 hours in that round. There will be no length or observing condition restrictions beyond this, but we will emphasize that programs that can tolerate sub-optimal conditions are more likely to be completed. If {\em two} of the provisionally accepted programs requested (say) 4 hours of IQ20 time, one would need to be removed. The Fast Turnaround Team would simply replace the program with the next-highest-ranked one roughly compatible with the available weather conditions.

To answer question (2), the team will perform technical assessments only on highly-ranked programs. These programs will also be checked for duplicate observations of targets defined in existing queue and classical programs. Programs with technical problems, or for which technical feasibility has not been adequately demonstrated, will be rejected outright and replaced with the next-highest-ranked program that fits in the available weather conditions. The short timescales involved in the program do not leave room for iteration with the PI in case of technical issues. 
To avoid accepting substandard proposals, an artificial oversubscription factor will be imposed if necessary, so that the total available hours or 30\% of the requested hours are scheduled each month, whichever is the smaller. Observations must be prepared within 10 days of proposals being accepted, and the Fast Turnaround Team will be available to work with PIs to ensure that this is possible.

\subsection{Observation Execution}
\label{exec}

Gemini is primarily a queue-scheduled observatory, and Fast Turnaround observations will be performed in this mode. However, rather than merging the observations into the regular queue, we plan to set aside distinct FT queue nights. The scheduling of these nights (three per month, divided between dark, grey and bright time) will be done at the start of each semester. Scheduling the Fast Turnaround observations on separate nights will make the program easier to track and monitor (very important during the initial trial period), avoid the need to take existing queue progress into account when soliciting Fast Turnaround proposals, simplify the decisions (scheduling priorities, instrument configuration changes, etc.) that must be taken by the queue coordinators, and allow the Fast Turnaround Team to better plan their time. Some flexibility will be permitted, such as executing time-critical Fast Turnaround programs on regular queue nights, and using normal queue observations to fill gaps on Fast Turnaround nights.

Observations will remain available for scheduling for three months from the observation preparation deadline, and will then be deactivated if not completed. To encourage rapid publication, the proprietary period of Fast Turnaround data will be limited to 6 months.

\section{PROPOSAL REVIEW MECHANISM}
\label{peer_review}

The details of Gemini's ``standard'' proposal selection process differ between the partners. Essentially,
however, the process involves a scientific review of proposals by one or more panels made up of astronomers serving terms of a few semesters.
Proposals are assessed by individual panel members who then confer, in
person or virtually, and come up with a final ranking to be merged with those of the other partners by the
International Time Allocation Committee. The proposals may or may not be sent to outside experts for
review, and assessments of technical feasibility are done in parallel with the TAC process.
This standard TAC process is familiar to Gemini's users, many of whom have served on TACs themselves.
Some of the perceived advantages of this system are that (1) all proposals in a given area are reviewed by the same group of people; (2) panelists read enough proposals to be able to gain a good sense of their relative merits; (3) large enough panels are likely to include at least one person knowledgeable about most proposals' subject areas; and (4) the panel discussion often results in a consensus about the final grades, which can differ significantly from those assigned before the group discussion. 

In principle there is no particular reason why the Fast Turnaround program would choose not to use some
version of this standard TAC process. In practice, though, we need to consider the possible effects of monthly
deadlines and the expected number of proposals. First, arranging monthly telecons between TAC members
seems impractical; these are, after all, people who volunteer their time for no immediate, personal reward (not
to mention the differences in time zones, teaching and other commitments, etc.). It is likely that any TAC
process involved in the Fast Turnaround Program would have to forego the panel discussion element.
With regard to the number of proposals, assuming an average proposal length of 5 hours, an oversubscription factor of 3, and 10\% of the science time at one telescope allocated to the program, suggests that the TAC would need to read about 15 proposals per month. While this is probably manageable during the initial trial, particularly if the TAC were split into Galactic and Extragalactic panels, it may be difficult to expand the program to both Gemini telescopes, or to use a larger fraction of the telescope time. 

We have decided instead to test an alternative:
``distributed peer review''. In this model, submitting a proposal
commits the submitter (or a delegate) to review other proposals.
Failure to complete the reviews results in the assessor's own proposal being
removed from consideration, and incentives can be built into the system to
encourage honest behaviour (Merrifield \& Saari 2009\cite{Merrifield09}).
The advantage of this kind of system is that it
removes the need to rely on external
reviewers who contribute their time on a strictly voluntary basis.
Proposers will submit when they have time and attention, as well as interest, to give to the project, which may
encourage more rapid publication of results. The ``cost'' of submitting a proposal
may discourage submission of weaker ones, and the system can also provide a valuable opportunity for
younger astronomers to gain insight into what makes a good proposal. 

This is the model that will be adopted for the Fast Turnaround program. We have raised the subject at several venues -- including university astronomy departments, national astronomy meetings, and the various Gemini oversight committees --and the responses have been quite illuminating. The
collaborative, community-based nature of distributed peer review appears to appeal to a significant segment
of Gemini's user community, while others have raised questions about the quality of the reviews and the possibility of abusing the system. There is often not a clear consensus about the potential dangers and how they should be mitigated. For instance, some have suggested that reviewers' grades should be weighted by their self-assessment of their knowledge of that subject area, with the opinion of self-declared experts given extra weight. Others, however, suggest that people will review proposals in their own field more critically than is necessarily warranted. Some are in favor of down-grading the proposals of the reviewers whose grades differ the most from those of the other reviewers, and may therefore be trying to penalize competing proposals, while others warn that this may unfairly penalize genuine differences of opinion.

Concerns about disclosure of ideas will be dealt with by requiring reviewers to declare conflicts of interest and agree to keep proposals confidential and use them only for the intended purpose. Those not comfortable with this approach will still be able to apply through the standard TAC route for the foreseeable future. Maintaining a healthy subscription rate will be important in dealing with some concerns about the quality of the proposals and their assessments. More proposals means a larger pool of reviewers, and a better chance that one's proposal will be reviewed by someone knowledgeable about the subject area. If the oversubscription rate is fairly high, the likelihood of weak proposals being accepted is reduced. Strong community interest in the Fast Turnaround scheme will be an important factor in its success.

Distributed peer review is new and unfamiliar and poses some interesting
challenges. With a careful design, and clear communication with the user community, however,
these issues do not seem insurmountable. Our approach will be to expect that people will behave ethically and responsibly, while keeping a careful watch for evidence that this is not the case. An advantage of the regular monthly deadlines is that the model can be continuously adapted as we gain experience with the program.

\subsection{Distributed Peer Review: A Field Experiment}

After the regular Gemini 2014B proposal deadline at the end of March 2014, PIs of Canadian-led proposals were asked by the Canadian National Gemini Office whether they would be willing to take part in a trial of the distributed peer review system. Participating astronomers would review each other's proposals, and the top-ranked proposal from this process would be guaranteed to have up to 10 hours scheduled in band 1 with rollover status (the ability to be carried over into future semesters), regardless of how it fared with the regular TAC. This would allow us to compare the peer-review ranking with that of the regular TAC; test the software developed to handle the peer review process; and gather feedback from the PIs about their experience with the process. Rapid scheduling and execution were not part of this trial.

Of the 29 PIs who were contacted, 9 agreed to take part. Their proposals spanned a wide range of scientific areas, from brown dwarfs and exoplanets to galaxies and supernovae in the distant Universe. A handful of PIs responded but declined the invitation. Their reasons were: that they would like to but were busy finishing their PhD thesis; that they felt that 10 proposals would be too many to review; and that they believed\footnotemark\ their proposal would be awarded band 1 time in any case. One of the 9 astronomers who agreed to take part did not actually do so. The grades received for this proposal were used for statistical purposes in this trial, but would have been excluded from the final ranking if this were not simply an experiment.

\footnotetext{Erroneously, as it turned out.}

The proposal handling software uses the keywords in the proposals to attempt to match proposals to reviewers knowledgeable about that subject area, but with the small number of participants in this trial, all PIs had to review all other proposals. The PIs were sent the titles and investigator lists of the proposals selected for them to review, and they first had to declare of conflicts of interest and accept the terms and conditions of the program (i.e., to keep the proposals confidential and use them only for the purpose of the review). Ten conflicts of interest were declared, and the reviewers often left detailed notes explaining the reason for the conflict. In most (8/10) cases the conflicts were ``symmetrical'', that is, if PI\#1 declared a conflict with PI\#2, the reverse was also true. In the other two cases, the conflict was because the reviewer had links with one of the co-Is of the other proposal. The symmetrical conflicts were generally based on the proposal PIs knowing each other, and/or the proposals obviously addressing very similar science or observing the same target(s). One conflict arose because of a suspicion that two proposals may be for the same object, although this was not clear from the proposal title. We are implementing automatic checking of target duplication based on the coordinates in the proposals, so that such cases may be avoided in the future.

On receipt of the completed conflict/terms forms, the software automatically sends out links to the full proposal files and review forms. The reviewers are asked to give each proposal a score from 1 (strongest) to 5 (weakest), and are asked to use the full scale and avoid extreme distributions (e.g. 1 high score and 9 low scores). They are also asked to rate their knowledge of the proposal's subject area, from ``1: I work or have recently worked in this area'' to ``3: I have little knowledge of this field''. The review forms also include a ``fatal flaw'' checkbox, available for flagging a proposal that has grave problems. This is currently intended to be used only for monitoring purposes, although it could eventually be used to exclude proposals from the final list. All but the above-mentioned PI returned their forms before the deadline. In this small-scale trial, the final ranked list was created by simply averaging the individual grades. However, the software also contains the capability to clip the highest and lowest scores before generating the final ranking.

\begin{figure}
   \begin{center}
   \begin{tabular}{c}
     \includegraphics[height=10cm]{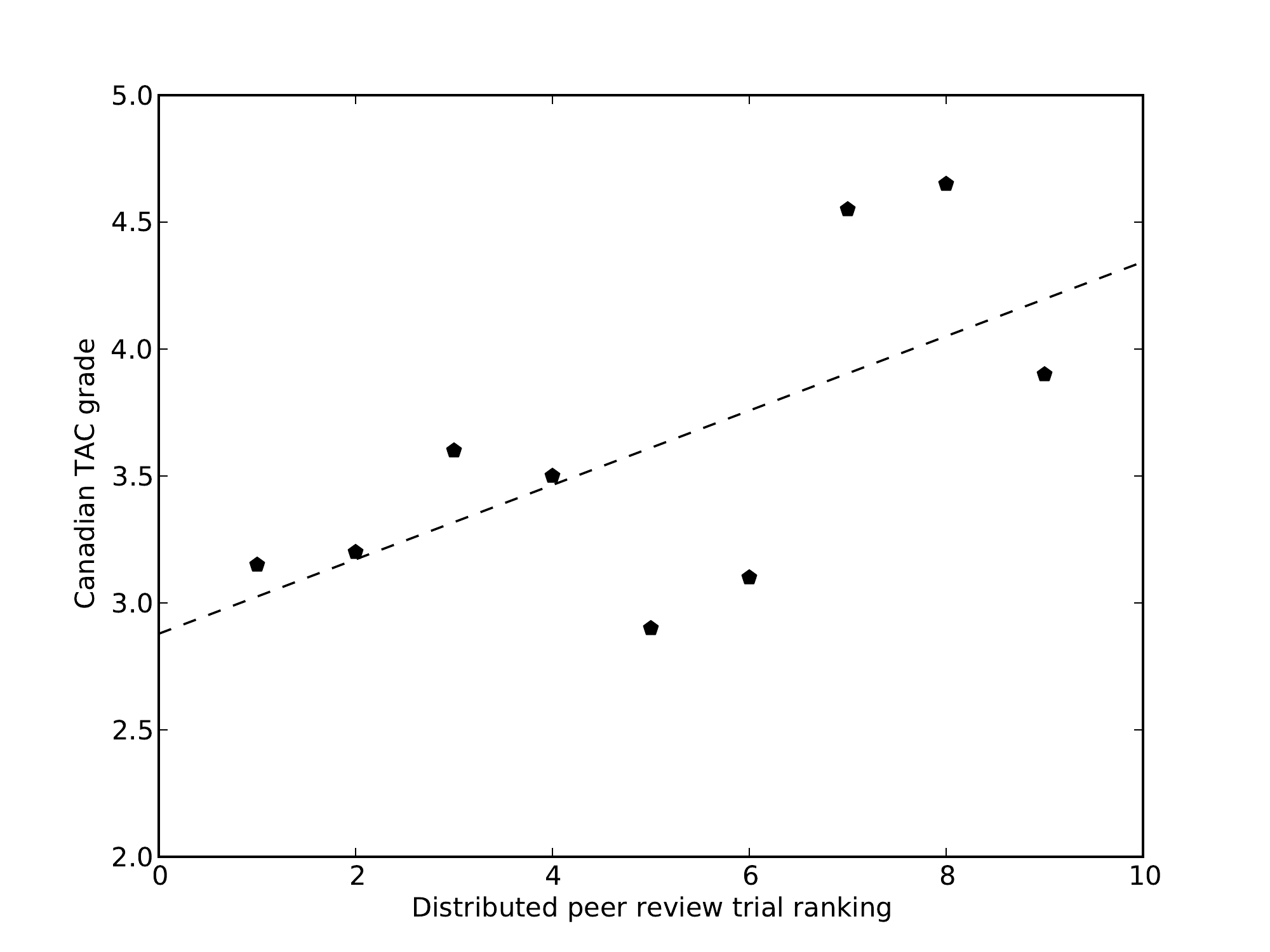}
   \end{tabular}
   \end{center}
   \caption{\label{compare} Comparison of proposal grades returned by the Canadian TAC for the 9 participating proposals with the final ranking from the distributed peer review trial. The CanTAC grades range from 1 (strongest) to 5 (weakest). }
   \end{figure}

Figure \ref{compare} shows that there is some correlation between the final peer review ranking and the grades of the TAC. This is perhaps surprising, given the small number of participants and the variety of scientific areas covered. PIs did not know in advance that they may be taking part in this trial, and therefore did not necessarily write for a general audience (regular Canadian Gemini proposals are sent to two expert reviewers whose comments are then available to the relevant galactic or extragalactic TAC). The two proposals that were ranked highest by the volunteers came third and fourth in the CanTAC ranking of these 9 proposals, and the three lowest-ranked proposals in the trial were also at the bottom of the TAC rankings. The programs that were rated highest by the TAC did not take part in this trial, but the agreement between the peer review and TAC opinion about the lowest-ranked proposals seems consistent with the received wisdom that the strongest and weakest proposals are easily identifiable. The ``winning'' program, a postdoc-led project using the Gemini Planet Imager, was placed in band 2 by the regular TAC.

After the review forms had been received, but before the TAC meeting, the participants were requested to give feedback about the process. They were asked about several areas, such as the length of time spent on reviewing the proposals, their opinion of the grading scheme, and what would make them more likely to participate in the program in the future. They were also asked whether their reviews could be (anonymously) returned to the relevant PIs. This will be the default when the Fast Turnaround pilot is launched, but individual permission was requested in this trial because  the relatively small number of people involved meant that guessing the identity of the reviewers would not be difficult.

Respondents reported taking from 45 minutes to review three proposals, to six hours to review eight proposals\footnotemark, with PhD students and postdocs taking significantly more time than tenured and tenure-track PIs. The word that came up most often in describing the reviewing was ``challenging'', followed by ``interesting''. Occasionally these were followed by ``time-consuming'' or ``painful''. The comments from PhD students and postdocs suggested that this process will be valuable to early-career researchers:

\footnotetext{Normally if a reviewer declares a conflict, they would be automatically assigned a replacement proposal. This was not possible with the small group taking part in the trial, leading to unequal numbers of proposals per reviewer.}

{\em ÒI found this challenging and time-consuming but actually EXTREMELY helpful. Since I cannot fill in the knowledge blanks with the proposals, reviewing these helped me better understand how to craft an easily understandable proposal.Ó

ÒI found it challenging, but definitely interesting. It felt like a very good exercise.Ó

ÒI found it both interesting and challenging. It was actually very useful for learning what's going on in these fields, more so than a journal club or astro-ph discussion.Ó}

Encouragingly, no participant was more than ``mildly concerned'' about how a ``devious'' PI might try to take advantage of the peer review system. Most (6/7 respondents) were against the idea of downgrading a PI's own proposal if their rankings differed significantly from the mean of the group, citing concerns that first-time reviewers would be unfairly penalised, as well as those with genuine differences of opinion. 

All respondents expressed interest in taking part in the Fast Turnaround program when it is launched at Gemini, from four who would take part ``for sure'', to two who were ``likely'' to participate. Receiving fewer proposals, or proposals in fields of study more closely related to the reviewer's, would make the participants more likely to take part, although one was against restricting the field of study because that would increase the chances of conflicts of interest. One exoplanet PI noted that the Fast Turnaround program will be useful for getting quickly data on candidates of high potential interest that soon set for the season.

In other comments, one PI noted some concern about the quality of the peer-review since the proposals will tend to be reviewed by non-specialists. However, as more proposals are received, the keyword-matching algorithm in the software, which attempts to assign proposals to reviewers in similar fields, should result in proposals and reviewers being somewhat better matched. Also, PIs may tend to write proposals in a way that makes them more compelling to a non-specialist audience. Others left positive remarks: {\em ``This was very helpful, thanks!'', ``It was painless'', ``I think it went well'', ``I participated because I felt I should help this process''.} While the sample was probably biased towards astronomers already interested in the Fast Turnaround program and comfortable with the idea of distributed peer review, the overall feedback was very encouraging.

\section{PROJECT STATUS AND IMPLEMENTATION PLAN}

The Fast Turnaround project was reviewed by a panel of internal and external experts (including representatives from the Gemini Users' Committee, Gemini Science and Technology Advisory Committee, Space Telescope Science Institute, and European Southern Observatory) in April 2014, and their recommendations incorporated in the program design. The following month, the Gemini Board of Directors approved the use of up to 10\% of the science time on Gemini North for a trial of the program. We plan to launch the trial in January 2015, initially for a six-month period but with the possibility of continuing if warranted by the demand for the program and the results it produces. The remainder of this year will be spent updating the supporting software, training the Fast Turnaround team, publicising the mode, and putting in place the mechanisms necessary to operate and monitor the system.

Monitoring and communication will be a fundamental part of the program. During the trial, the support staff will make regular web updates with basic information about the accepted programs. They will also write a summary of each Fast Turnaround night, letting the community and other stakeholders know which programs were started, attempted, or completed, and whether time was lost to weather and other issues. More detailed information will also be collected, analysed, and disseminated, such as:

\begin{itemize}
\item Feedback from users, including those whose proposals were not accepted; 
\item The monthly oversubscription factor of the program, and number of proposals received; 
\item Statistics such as the dispersion in the grades of each proposal; 
\item Relationships, if any, between e.g. reviewer's self-assessment and proposal grade; 
\item Proposal success rates by science category, PI gender, hours requested, etc.; 
\item Proposals rejected because of missed deadlines, technical problems, etc.; 
\item Interventions made by support staff (e.g. disallowing a complex observing mode that could not be
supported); 
\item Program completion rates; 
\item Impact on regular observing programs, if applicable; 
\item Status of the data (useful, reduced, published, ...); 
\item Use of data (e.g. completes a larger data set, pilot observation for longer program, etc.); 
\item Partner time usage
\end{itemize}

The program has been designed such that, if it  is successful and the demand for time is significant, there should be no barriers to expanding the mode to operate at Gemini South and/or to take a larger share of the telescope time.

\section{CONCLUSIONS}

While monthly proposal deadlines have been operated at other observatories (e.g. the UKIRTSERVICE program\cite{Howat96}), and distributed peer review is being trialled for certain National Science Foundation grant programs\footnotemark, to the best of our knowledge the Fast Turnaround program will be the first instance of these concepts being combined in astronomy. Through this novel observing mode, Gemini aims to become increasingly responsive to its users and to offer a wider variety of ways to achieve scientific objectives. The interest and support of the astronomical community will be key to making the program a success, and potential users are welcome to contact us with their questions and comments.

\footnotetext{www.nsf.gov/pubs/2013/nsf13096/nsf13096.pdf}

\acknowledgments     

We are grateful to the Fast Turnaround review committee for providing a thorough assessment of the program and many thoughtful suggestions. The panel members were: M. Rejkuba (chair), A. Fruchter, S. Leggett, S. Margheim, A. Rest, and T. Storchi-Bergmann. We also appreciate the members of the Canadian Gemini community who took part in the distributed peer review trial, and I. N. Reid and G. Hazelrigg for helpful discussions about the proposal review process. This work was supported by the Gemini Observatory, which is operated by the Association of Universities for Research in Astronomy, Inc., on behalf of the international Gemini partnership of 
Argentina, Australia, Brazil, Canada, Chile, and the United States of America.


\bibliography{report}   
\bibliographystyle{spiebib}   

\end{document}